\documentclass[11pt,eqs]{article}
\usepackage{latexsym,amsmath}

\def\cleq{\setcounter{equation}{0}}
\title{Twisted C-brackets
\thanks{Work supported in part by
Institute of Physics, Belgrade and Serbian Ministry of Science and Technological Development }}
\author{ Lj. Davidovi\'c \thanks{e-mail: ljubica@ipb.ac.rs}, I. Ivani\v sevi\'c \thanks{e-mail: ivanisevic@ipb.ac.rs} and B. Sazdovi\'c
\thanks{e-mail: sazdovic@ipb.ac.rs}\\
{\it Institute of Physics, University of Belgrade}\\
{\it Pregrevica 118, 11080 Belgrade, Serbia}}

\begin{document}
\maketitle
\begin{abstract}
We consider the double field formulation of the closed bosonic string theory, and calculate the Poisson bracket algebra of the symmetry generators governing both general coordinate and local gauge transformations. Parameters of both of these symmetries depend on a double coordinate, defined as a direct sum of the initial and T-dual coordinate. When no antisymmetric field is present, the $C$-bracket appears as the Lie bracket generalization in a double theory. With the introduction of the Kalb-Ramond field, the $B$-twisted $C$-bracket appears, while with the introduction of the non-commutativity parameter, the $\theta$-twisted $C$-bracket appears. We present the derivation of these brackets and comment on their relations to analogous twisted Courant brackets and T-duality.
\end{abstract}

\section{Introduction}
\cleq

The development of string theory led to the discovery of T-duality \cite{ODD, tdual, tdual1}, a transformation needed to establish a connection between the existing string theories. In case of closed strings, the emergence of T-duality is closely related to the fact that they can wind around compact dimension. Two theories defined in different geometries can create the same string spectrum, and are said to be T-dual. The T-duality was generalized into procedures of finding mutually physically equivalent descriptions of a string \cite{buscher, wcb}. 

The T-duality opened a search for the connections between the relevant mathematical structures within T-dual theories. Some of them appear upon consideration of the generators of symmetries, which in the classical theory act on the energy-momentum tensor via Poisson bracket. The changes of energy-momentum tensor under the symmetry transformations can be interpreted as the changes in the space-time fields under diffeomorphisms and local gauge transformations. Under these transformations, the Virasoro algebra is not broken, and the corresponding conformal field theories of the string are isomorphic \cite{evans1, evans2}. In the Poisson bracket algebra of symmetry generator governing both diffeomorphisms and local gauge transformations the Courant bracket \cite{courant, courant1} is obtained. It is known that under T-duality the bosonic string symmetries transform into one another \cite{dualsim}, and as such, the Courant bracket represents the self T-dual extension of the Lie bracket. 

The Courant bracket and its various deformations have found their appearances in string theory in multiple occasions. For instance, the relevant string theory fluxes \cite{stw} appeared in the twisted Courant brackets \cite{twist}, that were obtained from the world-sheet \cite{c, nick1, nick2} and in the algebroid relations of the appropriate vielbeins \cite{flux1, royt, flux-bianchi}. They can be obtained in the algebra of symmetry generators as well, provided the generator is expressed in the appropriate non-canonical basis. These non-canonical variables, also known as currents, give rise to the string fluxes as its structure constants in their Poisson bracket algebra. In the presence of Kalb-Ramond field, the $B$-twisted Courant bracket was obtained in the generator algebra \cite{csd}. In the self T-dual picture, obtained by swapping the canonical momenta with coordinate $\sigma$-derivatives, and the background fields with their T-duals, the algebra bracket becomes the $\theta$-twisted Courant bracket \cite{csd}. The former twisted Courant bracket contains geometric $H$-flux, while the latter contains non-geometric $Q$ and $R$ fluxes. It was showed that they are mutually related by T-duality \cite{crdual}. The Courant bracket twisted simultaneously by both $B$ and $\theta$ has also been constructed from the symmetry generator algebra \cite{thbtwist}. The bracket contains all string fluxes and is invariant under T-duality. 

The observations regarding T-duality interchanging different twisted Courant brackets were done in the single theory. However, there is a growing interest in the double theory approach, in which all background fields are function both of the initial coordinate and the T-dual coordinate. The T-duality is a symmetry of a double action, and can be simply realized by permutation of the coordinates \cite{perm2, perm1, BS, BS1}. We would like to see how the interchange of different twisted Courant brackets and their fluxes under T-duality is manifested in the double field theory.

The first steps towards the double field theory development were done in the early 1990s \cite{DFT-1, DFT-2, DFT-3, DFT-4} (for nice reviews on the subject, see \cite{DFTreview, DFTreview1}). In \cite{siegel1, siegel}, generalized Lie derivative was constructed in the double space, which acting on another vector defined in the double space gives rise to what is now known as the $C$-bracket. It is an extension of the Lie bracket to the double space, and found important role in governing symmetries algebra in double field theory. The action on the double torus invariant under both initial and T-dual diffeomorphisms was formulated in \cite{DFT1}, where background fields were expanded around constant values up to the cubic order. It was later generalized to a generalized metric formulation of the double theory \cite{DFT3}. In \cite{DFT2}, it was demonstrated that the $C$-bracket reduces to the Courant bracket, when dependence on T-dual coordinates is neglected. From this correspondence between $C$- and Courant bracket we expect that the twisted Courant brackets are related to some twisted $C$-bracket, which we consider in this paper by analyzing the Poisson bracket algebra of appropriate generators.  


We begin by outlining the properties of a closed bosonic string in the double space, when only metric tensor is present. We consider the standard Poisson bracket relations within the initial and T-dual phase space, while the algebra of the variables from different phase spaces is obtained from the requirement that the T-duality transformations commute with the Poisson bracket relations. The algebra of both generalized coordinate transformations and their T-duals in this background was already considered in \cite{csd}, where it was showed that it produces the $C$-bracket. We include this case in this paper for completeness and as a prerequisite for future chapters. 

Afterwards, we outline a simple procedure of obtaining the twisted Courant bracket from the generator written in a suitable basis, introduced in \cite{thbtwist}. We demonstrate that this procedure is applicable to the double theory as well, and that it can be used to define new brackets in double theory. By analogy with how twisted Courant brackets are defined, we define these brackets as the twisted $C$-brackets. From a differential graded algebra perspective the twisted $C$-brackets were briefly discussed in \cite{deser}. 

Subsequently, we introduce the Kalb-Ramond field dependent on both coordinates $x^\mu$ and $y_\mu$ to the metric only background. We show that the usual expressions for the Lagrangian and the Hamiltonian are obtained, when $B$-transformations act on a diagonal generalized metric. The same transformation can be seen as keeping the generalized metric diagonal, while transforming the basis. The transformed basis includes non-canonical momenta, the Poisson bracket of which includes the double theory flux. We calculate the Poisson bracket algebra of this new generator, and obtain what in our notation corresponds to the $B$-twisted $C$-bracket. 

We consider the background characterized with the T-dual metric tensor only, and introduce the non-commutative parameter with the action of  $\theta$-transformations. The Hamiltonian once again can be expressed in a diagonal form, in the basis spanned by non-canonical momenta. In their Poisson bracket algebra, we show that non-canonical momenta feature $\hat{\Theta}$ double flux. We express the symmetry generator in this basis, and obtain the $\theta$-twisted $C$-bracket in their algebra governed by Poisson bracket. By construction, this bracket is T-dual to the $B$-twisted $C$-bracket.

We note that both the initial theory, in which all variables depend solely on $x^\mu$, and the T-dual theory, in which all variables depend solely on $y_\mu$ can be obtained from the double theory, by demanding that there is no dependence on $y_\mu$ in the former, and no dependence on $x^\mu$ in the latter case. In the initial theory, the $B$-twisted $C$-bracket reduces to the $B$-twisted Courant bracket, while the $\theta$-twisted $C$-bracket becomes the $\theta$-twisted Courant bracket. On the other hand, in the T-dual theory the $B$-twisted $C$-bracket reduces to the $\theta$-twisted Courant bracket, while the $\theta$-twisted $C$-bracket becomes the $B$-twisted Courant bracket. This way, we show that projections of the twisted $C$-brackets to mutually T-dual phase spaces establish the twisted Courant brackets that are mutually T-dual \cite{crdual}.

\section{Bosonic string in a double metric space}
\cleq
Consider the closed bosonic string in the background defined solely by the coordinate dependent metric field $G_{\mu \nu}(X)$. The metric tensor is a function of a double coordinate $X^M$, defined in a direct sum of the initial coordinate space, characterized by $x^\mu$, and T-dual coordinate space, characterized by $y_\mu$
\begin{equation}
X^M = \begin{pmatrix}
x^\mu \\
y_\mu \\
\end{pmatrix} \, ,
\end{equation}
where $\mu = 0, 1 ... D -1, \ D= 26$. In the conformal gauge, the Lagrangian density is given by \cite{action, regal}
\begin{equation}\label{eq:action}
{\cal{L}} = \frac{\kappa }{2} \partial_+ X^M G_{MN} \partial_- X^N \,  ,\ \partial_{\pm} X^M = \dot{X}^M \pm X^{\prime N} \, ,
\end{equation}
where 
\begin{equation} \label{eq:GMN}
G_{MN} = 
\begin{pmatrix}
G_{\mu \nu} & 0 \\
0 & (G^{-1})^{\mu \nu}
\end{pmatrix} \, .
\end{equation}
The canonical momenta are obtained from the variation of the Lagrangian 
\begin{equation}
\Pi_M = \kappa G_{MN} \dot{X}^N \equiv \begin{pmatrix}
\pi_\mu \\
{}^\star \pi^\mu
\end{pmatrix} \, , \quad \pi_\mu = \kappa G_{\mu \nu} \dot{x}^\nu \, , \ {}^\star \pi^\mu = \kappa (G^{-1})^{\mu \nu} \dot{y}_\nu \, ,
\end{equation}
while the Legendre transformation of the Lagrangian gives the canonical Hamiltonian
\begin{equation} \label{eq:HG}
{\cal H}_{C} = \frac{1}{2\kappa} \Pi_M G^{MN} \Pi_N + \frac{\kappa}{2} X^{\prime M} G_{MN} X^{\prime N} \, ,
\end{equation}
where 
\begin{equation}
G^{MN} = \begin{pmatrix}
(G^{-1})^{\mu \nu} & 0 \\
0 & G_{\mu \nu} 
\end{pmatrix} \, .
\end{equation}
The indices on $G$ are lowered by the $O(D,D)$ invariant metric, given by
\begin{equation}
\eta_{MN} = 
\begin{pmatrix}
0 & 1 \\
1 & 0 
\end{pmatrix} \, ,
\end{equation}
i.e. $G^{MN} = \eta^{MP} \eta^{NQ} G_{PQ}$.

In the conventional field theory, the transformation of the metric tensor under diffeomorphisms parametrized with $\xi$ is governed by the Lie derivative $\delta_\xi G_{\mu \nu} = {\cal L}_{\xi} G_{\mu \nu}$, and generated by $\xi^\mu \pi_\mu$. The Poisson bracket algebra of such a generator gives rise to the Lie bracket.

Analogously, one defines the diffeomorphisms in the double space by 
\begin{equation}
G(\Lambda) = \int d\sigma {\cal G}_\Lambda  \, ,
\end{equation}
with
\begin{equation}\label{eq:Gen}
 {\cal G}_\Lambda  =  (\Lambda^T)^M (X) \eta_{M N} \Pi^N  \, \qquad \Longleftrightarrow \qquad {\cal G}_\Lambda  = \langle \Lambda, \Pi   \rangle \,  ,
\end{equation}
where $\Lambda^M$ are the parameters, given by
\begin{equation}
\Lambda^M  (X) = 
\begin{pmatrix}
\xi^\mu  (x, y) \\
\lambda_\mu (x, y) 
\end{pmatrix} \, ,
\end{equation}
and $\langle, \rangle$ is the natural inner product in a double space
\begin{equation} \label{eq:skalproizvod}
\langle \Lambda, X \rangle = (\Lambda^T)^M \eta_{MN} X^N \, .
\end{equation}

To give an interpretation to the generator (\ref{eq:Gen}), we expand the inner product and obtain 
\begin{equation}
{\cal G}_\Lambda = \xi^\mu (x, y) \pi_\mu + \lambda_\mu (x, y)  {}^\star \pi^\mu \, .
\end{equation}
The first term represents the diffeomorphisms, while the second term represents the local gauge transformations, equivalent to the T-dual diffeomorphisms \cite{dualsim}. The algebra of this generator leads to the double theory generalization of the Lie bracket. 

\subsection{$C$-bracket}

To obtain the Poisson bracket algebra of the generator ${\cal G}_\Lambda$, one uses the Poisson bracket relations between the double variables (for details, see \cite{csd})
\begin{eqnarray} \label{eq:PBPX}
\{ \Pi_M (\sigma), \Pi_N ({\bar \sigma}) \} &\simeq& \kappa \, \eta_{M N} \delta^\prime (\sigma - {\bar \sigma} ) \, , \\ \notag
\{ X^{\prime M} (\sigma), \Pi_N (\bar{\sigma}) \} &=&  \delta^M_{\ N}\ \delta^\prime (\sigma - {\bar \sigma} ) \, ,
\end{eqnarray}
the same relations as used in \cite{siegel1, siegel}. The terms stemming from bracket between only coordinates $\{ X^M (\sigma), X^N (\bar{\sigma}) \}$ are canceled by the appropriate choice of Heaviside theta function, which sets all parameters and fields annihilated by the operator 
\begin{equation} \label{eq:Dcon}
\eta^{MN} \partial_M \partial_N = \partial^M \partial_M =  0 \, ,
\end{equation}
where $\partial^M$ are the derivatives in a double theory, given by
 \begin{eqnarray}
\partial_M =  \begin{pmatrix}
  \partial_\mu  \\
  \tilde{\partial}^\mu  \\
 \end{pmatrix}  \, ,   \qquad  \Big( \partial_\mu \equiv \frac{\partial}{\partial x^\mu} \, , \quad  \tilde{\partial}^\mu \equiv \frac{\partial}{\partial y_\mu}  \Big) \, .
\end{eqnarray}
Furthermore, we also require that the product of any two fields $\phi$ and $\psi$ is also annihilated by (\ref{eq:Dcon}), i.e.
\begin{equation} \label{eq:mix}
\partial^M \partial_M (\phi \psi) = (\partial^M \partial_M \phi )\ \psi+2\partial^M \phi \ \partial_M \psi+ \phi \ \partial^M \partial_M \psi= 2\partial^M \phi \ \partial_M \psi= 0 \, .
\end{equation}
The conditions (\ref{eq:Dcon}) and (\ref{eq:mix}) were also used in \cite{DFTreview, siegel1, siegel}, and are referred to as strong condition.

It follows that 
\begin{equation} \label{eq:GGCbr}
\{{\cal G}_{\Lambda_1} (\sigma), {\cal G}_{\Lambda_2}(\bar{\sigma}) \} = - {\cal G}_{[\Lambda_1, \Lambda_2]_{\textbf C} } (\sigma)   \delta (\sigma - {\bar \sigma}) +\frac{\kappa}{2} \Big(\langle \Lambda_1, \Lambda_2 \rangle (\sigma) + \langle \Lambda_1, \Lambda_2 \rangle (\bar{\sigma}) \Big) \delta^{\prime}(\sigma-\bar{\sigma}) \, .
\end{equation}
where $[\Lambda_1, \Lambda_2]_C$ stands for the $C$-bracket, firstly defined by Siegel \cite{siegel1, siegel}. It was defined by
\begin{eqnarray}\label{eq:Cbrdef}
{[\Lambda_1, \Lambda_2]_{\textbf C}}^Q =  \Lambda_1^N \partial_N \Lambda_2^Q   -  \Lambda_2^N \partial_N \Lambda_1^Q    -
\frac{1}{2}  \eta_{NP}  \,  \Big( \Lambda_1^N \partial^Q \Lambda^P_{2}    - \Lambda_2^N \partial^Q \Lambda^P_{1}  \Big)    \, .
\end{eqnarray}

It is a well known fact that if the parameters $\Lambda$ do not depend on the T-dual coordinates $y_\mu$, the $C$-bracket reduces to the Courant bracket \cite{courant}
\begin{equation}
[ \Lambda_1, \Lambda_2 ]_{ {\textbf C}} \to  [ \Lambda_1, \Lambda_2 ]_{\cal{C}} =  \Lambda \equiv \begin{pmatrix}
\xi \\
\lambda
\end{pmatrix} \,  ,
\end{equation}
where
\begin{eqnarray} \label{eq:courantt}
\xi^\mu &=& \xi_1^\nu \partial_\nu \xi_2^\mu - \xi_2^\nu \partial_\nu \xi_1^\mu \, , \\ \notag
\lambda_\mu &=& \xi_1^\nu (\partial_\nu \lambda_{2 \mu} - \partial_\mu \lambda_{2 \nu}) - \xi_2^\nu (\partial_\nu \lambda_{1 \mu} - \partial_\mu \lambda_{1 \nu})+\frac{1}{2} \partial_\mu (\xi_1 \lambda_2- \xi_2 \lambda_1 ) \, .
\end{eqnarray}
The Courant bracket is defined on the generalized tangent bundle, that is a sum of the tangent and the cotangent bundle over a manifold. The elements of the generalized tangent bundle are the generalized vectors that combine vectors and 1-forms into a single entity \cite{gualtieri}.

We see that the symmetry generator governing general coordinate and local gauge transformations produces the $C$-bracket, in a same way that the generator of general coordinate transformations produces the Lie bracket. The $C$-bracket is therefore the T-dual invariant generalization of the Lie bracket in a double space, when there is no antisymmetric field. 

\subsection{$O(D,D)$ group}

At the end of this chapter, we outline the mathematical requirements that allows us to introduce the antisymmetric field to the double theory, and obtain the twisted $C$-bracket for any orthogonal transformation. Assume a transformation ${\cal O}$ that transforms momenta and gauge parameters as $\hat{\Pi} = {\cal O} \Pi$, and $\hat{\Lambda} = {\cal O} \Lambda$.  The generator (\ref{eq:Gen}) in terms of these new variables is given by
\begin{equation}
{\cal \hat{G}}_{\hat{\Lambda}} = \langle \hat{\Lambda}, \hat{\Pi}  \rangle = \langle {\cal O} \Lambda, {\cal O} \Pi  \rangle \, .
\end{equation}
If furthermore, we demand that the transformation is orthogonal \cite{ODD}, so that
\begin{equation} \label{eq:Oddtrans}
{\cal O}^T \eta {\cal O} = \eta \, ,
\end{equation}
the relation (\ref{eq:GGCbr}) becomes
\begin{eqnarray} \label{eq:OLOL1}
\Big\{ \langle \hat{\Lambda}_1, \hat{\Pi} \rangle (\sigma), \langle \hat{\Lambda}_2, \hat{\Pi} \rangle (\bar{\sigma})\Big\} &=& -\langle {\cal O} [{\cal O}^{-1} \hat{\Lambda}_1, {\cal O}^{-1} \hat{\Lambda}_2 ]_{{\textbf C}}, \hat{\Pi}\rangle  (\sigma) \delta(\sigma-\bar{\sigma})  \\ \notag
&&+\frac{\kappa}{2} \Big(\langle \hat{\Lambda}_1, \hat{\Lambda}_2 \rangle (\sigma) + \langle \hat{\Lambda}_1, \hat{\Lambda}_2 \rangle (\bar{\sigma}) \Big) \delta^{\prime}(\sigma-\bar{\sigma})  \, ,
\end{eqnarray}
the right hand side of which gives rise to the twisted $C$-bracket
\begin{equation} \label{eq:twistOdef}
[\hat{\Lambda}_1, \hat{\Lambda}_2]_{{\textbf C}_{\cal O}} = {\cal O} [{\cal O}^{-1} \hat{\Lambda}_1, {\cal O}^{-1} \hat{\Lambda}_2 ]_{{\textbf C}} \, .
\end{equation}
This is the twisted $C$-bracket, in a similar way that twisted Poisson, or twisted Courant brackets are defined.

We see that the every $O(D,D)$ transformation has a corresponding twisted $C$-bracket (\ref{eq:twistOdef}). It can be obtained in the Poisson bracket algebra of generator in the basis of new momenta $\hat{\Pi} = {\cal O} \Pi$. This procedure, that was applied for different twisted Courant brackets in \cite{csd, thbtwist}, will be used to obtain the $C$-bracket twisted by a 2-form, and by a bi-vector.

\section{Bosonic string in a complete double space}
\cleq

To obtain the action describing the string moving in doubled space described by both metric $G(X)$ and the antisymmetric Kalb-Ramond field $B(X)$, one considers the $B$-transformations
\begin{equation} \label{eq:eb}
(e^{\hat{B}})^M_{\ N} = \begin{pmatrix}
\delta^\mu_\nu & 0 \\
2B_{\mu \nu} & \delta^\nu_\mu
\end{pmatrix} \, ,
\end{equation}
where
\begin{equation} \label{eq:bhat}
\hat{B}^M_{\ N} = 
\begin{pmatrix}
0 & 0 \\
2B_{\mu \nu} & 0 \\ 
\end{pmatrix} \, .
\end{equation}
One easily finds 
\begin{equation} \label{eq:ebminus}
( (e^{\hat{B}})^T)_M^{\ N} = \begin{pmatrix}
\delta^\nu_\mu& -2B_{\mu \nu} \\
0 & \delta^\mu_\nu 
\end{pmatrix} \, ,
\end{equation}
and that the $B$-transformations are $O(D,D)$ (\ref{eq:Oddtrans}), i.e.
\begin{equation} \label{eq:eOe}
((e^{\hat{B}})^T)_M^{\ K} \ \eta_{KL} \ (e^{\hat{B}})^L_{\ N} = \eta_{MN} \, .
\end{equation}

The doubled action is governed by the generalized metric $H_{MN}$, the $B$-transformation of the doubled metric $G_{MN}$, defined in (\ref{eq:GMN})
\begin{equation} \label{eq:genmet}
H_{MN} = ( (e^{\hat{B}})^T)_M^{\ K}\ G_{KL}\ (e^{\hat{B}})^L_{\ N}  = 
\begin{pmatrix}
G^E_{\mu \nu} & - 2B_{\mu \rho} (G^{-1})^{\rho \nu} \\
2(G^{-1})^{\mu \rho} B_{\rho \nu} & (G^{-1})^{\mu \nu}
\end{pmatrix} \, ,
\end{equation}
where $G^E$ is the effective metric, given by
\begin{equation} \label{eq:Geff}
G^E_{\mu \nu} = G_{\mu \nu} - 4 (B G^{-1} B)_{\mu \nu} \, .
\end{equation}
The Lagrangian density for closed bosonic string is now given by 
\begin{eqnarray}\label{eq:DA}
{\cal L} = \frac{\kappa }{2} \partial_+ X^M   H_{M N}  \partial_-X^N \, .
\end{eqnarray}
This theory was considered in a constant and weakly curved background in \cite{BS,BS1}, where it was shown that this representation of doubled theory makes T-duality just a coordinate permutation, and represents all T-dual theories in an unified manner.

The equation of motion for (\ref{eq:DA}) is 
\begin{eqnarray}\label{eq:EMD}
\partial_+ ( H_{M N}  \partial_-  X^N) + \partial_- (H_{M N}  \partial_+ X^N) = 0 \, .
\end{eqnarray}

The conjugate momenta in a double space 
\begin{equation} \label{eq:PiM}
\Pi_M = \begin{pmatrix}
\pi_\mu \\
{}^\star \pi^\mu
\end{pmatrix} \, ,
\end{equation}
are as usual obtained by varying the Lagrangian density over $\dot{X}^M$
\begin{equation} 
\Pi_M =\kappa H_{MN} \dot{X}^N \, ,
\end{equation}
from which we have
\begin{equation} \label{eq:pi1}
\pi_\mu = G^E_{\mu \nu} \dot{x}^{\nu} - 2 (B G^{-1})_{\mu}^{\ \nu} \dot{y}_\nu \, ,
\end{equation}
and
\begin{equation} \label{eq:Tpi1}
{}^\star \pi^\mu =  (G^{-1})^{\mu \nu} \dot{y}_\nu + 2 (G^{-1} B)^{\mu}_{\ \nu} \dot{x}^{\nu} \, .
\end{equation}

The canonical Hamiltonian is given by 
\begin{equation} \label{eq:Hcan}
{\cal H}_{\cal C} = \frac{1}{2\kappa} \Pi_M  H^{MN} \Pi_N + \frac{\kappa}{2} X^{\prime M} H_{MN} X^{\prime N} \, .
\end{equation}

\subsection{T-duality}

The T-duality relations were firstly obtained by Buscher \cite{buscher} in case of existence of a global isometry. As the initial theory is described with the metric tensor $G_{\mu \nu}$ and the anti-symmetric Kalb-Ramond field $B_{\mu \nu}$, the T-dual theory is described with the T-dual metric tensor ${}^\star  G^{\mu \nu}$ and T-dual ${}^\star B^{\mu \nu}$ field, where
\begin{equation} \label{eq:GBstar}
{}^\star G^{\mu \nu} = (G_{E}^{-1})^{\mu\nu} \, , \quad {}^\star B^{\mu \nu} = \frac{\kappa}{2}\theta^{\mu \nu} \, ,
\end{equation}
and $\theta^{\mu \nu}$ is the non-commutativity parameter, given by
\begin{equation} \label{eq:thetadef}
\theta^{\mu \nu} = -\frac{2}{\kappa} (G^{-1}_E)^{\mu \rho}  B_{\rho \sigma} (G^{-1})^{\sigma \nu} \, .
\end{equation}

In a Lagrangian approach, relations between the coordinates of mutually T-dual phase spaces are given by 
\begin{eqnarray}\label{eq:xtdual}
\partial_{\pm}x^\mu \simeq
-\kappa\theta^{\mu\nu}_{\pm}  \partial_{\pm} y_\nu
\,  , \quad
\partial_{\pm}y_\mu\simeq
-2\Pi_{\mp\mu\nu} \partial_{\pm} x^\nu \, ,
\end{eqnarray}
where
\begin{eqnarray}
\Pi_{\pm\mu\nu} =
B_{\mu\nu} \pm\frac{1}{2}G_{\mu\nu} \, ,\quad  {\theta}^{\mu\nu}_{\pm}=
{\theta}^{\mu\nu}\mp \frac{1}{\kappa}(G_{E}^{-1})^{\mu\nu} \, .
\end{eqnarray}
In a double space approach, the above T-duality relations are combined so that they appear simply as a permutation of doubled coordinates \cite{perm2, perm1, BS, BS1}
\begin{equation} \label{eq:TDZ}
\partial_{\pm} X^{M} \simeq \pm \eta^{MN} H_{NK}  \,\partial_{\pm}X^K \, ,
\end{equation}
or equivalently as 
\begin{eqnarray}\label{eq:TD1}
\Pi_M  \simeq \kappa \, \eta_{M N}  X^{\prime M}  \, .
\end{eqnarray}
The consistency condition \cite{BS} for the T-dual transformation (\ref{eq:TDZ}) is actually the equation of motion of the doubled theory (\ref{eq:EMD}).

When the T-duality rules (\ref{eq:xtdual}) are applied to relations (\ref{eq:pi1}) and (\ref{eq:Tpi1}), one easily obtains the usual expressions for the initial theory momenta
\begin{equation} \label{eq:pidef}
\pi_\mu = \kappa G_{\mu \nu} \dot{x}^\nu - 2\kappa B_{\mu \nu} x^{\prime \nu} \, ,
\end{equation}
and similarly for the T-dual momenta 
\begin{equation} \label{eq:Tpidef}
{}^\star \pi^\mu = \kappa (G_E^{-1})^{\mu \nu} \dot{y}_\nu - \kappa^2 \theta^{\mu\nu} y^{\prime}_{\nu} \, .
\end{equation}

Lastly, let us comment on the fact that T-duality transforms the equations of motions into the Bianchi identities of the T-dual theory \cite{wcb}. In the double space formulation, the equations of motion and Bianchi identities are united \cite{BS} in a single relation (\ref{eq:EMD}).

\subsection{$C$-bracket twisted by $B$}

To obtain the $C$-bracket twisted by $B$, we rewrite the Hamiltonian (\ref{eq:Hcan}) in terms of diagonal $G_{MN}$
\begin{equation}
{\cal H}_{\cal C} = \frac{1}{2\kappa} \hat{\Pi}_M  G^{MN} \hat{\Pi}_N + \frac{\kappa}{2} \hat{X}^{\prime M} G_{MN} \hat{X}^{\prime N} \, ,
\end{equation}
where the new momenta are
\begin{eqnarray}\label{eq:PiHat}
 \hat{\Pi}^M  = (e^{\hat{B}})^M{}_N  \Pi^N   =
 \begin{pmatrix}
\delta^\mu_\nu & 0 \\
2B_{\mu \nu} & \delta^\nu_\mu
\end{pmatrix}
\begin{pmatrix}
{}^\star \pi^\nu     \\
\pi_\nu   \\
\end{pmatrix}
=
\begin{pmatrix}
{}^\star \pi^\mu     \\
\pi_\mu  + 2B_{\mu \nu} {}^\star \pi^\nu    \\
\end{pmatrix}
\equiv
\begin{pmatrix}
{}^\star \pi^\mu     \\
{\hat \pi}_\mu    \\
\end{pmatrix}
   \, ,
\end{eqnarray}
and new coordinates
\begin{equation}
 \hat{X}^{\prime M}  = (e^{\hat{B}})^M{}_N  X^{\prime N}   =
 \begin{pmatrix}
\delta^\mu_\nu & 0 \\
2B_{\mu \nu} & \delta^\nu_\mu
\end{pmatrix}
\begin{pmatrix}
x^{\prime \nu}     \\
y^{\prime}_\nu   \\
\end{pmatrix} = 
\begin{pmatrix}
x^{\prime \mu} \\
y_{\mu}^{\prime} + 2B_{\mu \nu} x^{\prime \nu}
\end{pmatrix} \equiv
\begin{pmatrix}
x^{\prime \mu} \\
\hat{y}_{\mu}^{\prime}
\end{pmatrix} 
   \, .
\end{equation}
We see that the $\sigma$-derivative $X^{\prime M}$ transforms as a vector under the $B$-shifts. The T-duality relations are preserved, i.e. 
\begin{equation}
\hat{\Pi}_M \simeq \kappa \eta_{MN} \hat{X}^{\prime N} \, .
\end{equation}

Let us now introduce new symmetry parameters
\begin{eqnarray}\label{eq:LHat}
 \hat{\Lambda}^M  = (e^{\hat{B}})^M{}_N\Lambda^N  =
  \begin{pmatrix}
\delta^\mu_\nu & 0 \\
2B_{\mu \nu} & \delta^\nu_\mu
\end{pmatrix}
\begin{pmatrix}
   \xi^\nu     \\
\lambda_\nu   \\
\end{pmatrix}
=
\begin{pmatrix}
\xi^\mu     \\
\lambda_\mu  + 2B_{\mu \nu}  \xi^\nu    \\
\end{pmatrix}
\equiv
\begin{pmatrix}
\xi^\mu     \\
 {\hat \lambda}_\mu   \\
\end{pmatrix} \, .
\end{eqnarray}
The symmetry generator can be written as
\begin{equation}
\hat{\cal G}_{\hat{\Lambda}}  = \hat{\Lambda}^M \eta_{MN} \hat{\Pi}^N \, ,
\end{equation}
and its algebra as
\begin{equation} \label{eq:algdef}
\{ \hat{\cal G} _{\hat{\Lambda}_1} (\sigma), \hat{\cal G}_{\hat{\Lambda}_2} (\bar{\sigma}) \} =  - \hat{\cal G}_{[ \hat{\Lambda}_1,  \hat{\Lambda}_2]_{{\textbf C}_B}} (\sigma) \delta(\sigma-\bar{\sigma}) +\frac{\kappa}{2} \Big(\langle \hat{\Lambda}_1, \hat{\Lambda}_2 \rangle (\sigma) + \langle \hat{\Lambda}_1, \hat{\Lambda}_2 \rangle (\bar{\sigma}) \Big) \delta^{\prime}(\sigma-\bar{\sigma}) \,  ,
\end{equation}
where 
\begin{equation} \label{eq:GGcourant34}
[ \hat{\Lambda}_1,  \hat{\Lambda}_2]_{{\textbf C}_B} =  e^{\hat{B}}  [e^{-\hat{B}} \hat{\Lambda}_1, e^{-\hat{B}}
 \hat{\Lambda}_2]_{{\textbf C}}  \, .
\end{equation}
This is a definition of the $B$-twisted C-bracket. To obtain its expression, let us firstly obtain the expressions for Poisson brackets between the new momenta and symmetry parameters. Using (\ref{eq:PiHat}), one obtains
\begin{eqnarray} \label{eq:hphp1}
\{ \hat{\Pi}^M (\sigma), \hat{\Pi}^N (\bar{\sigma}) \} &=& \{ (e^{\hat{B}} \Pi)^M  (\sigma), (e^{\hat{B}} \Pi)^N (\bar{\sigma}) \}  \\ \notag
 &=&(e^{\hat{B}} )^M_{\ J} (\sigma)(e^{\hat{B}} )^N_{\ K} (\bar{\sigma}) \{\Pi^J (\sigma), \Pi^K(\bar{\sigma}) \} \\ \notag
&&- (e^{\hat{B}})^M_{\ J}\partial^J (e^{\hat{B}})^N_{\ K} \Pi^K \delta(\sigma-\bar{\sigma}) + (e^{\hat{B}})^N_{\ J} \partial^J (e^{\hat{B}})^M_{\ K} \Pi^K \delta(\sigma-\bar{\sigma}) \, ,
\end{eqnarray}

where unless written otherwise, dependence on $\sigma$ is assumed. 
To obtain the first term, we apply the T-dual relations (\ref{eq:PBPX})
\begin{eqnarray}\label{eq:PBL3}
(e^{\hat{B}} )^M_{\ J} (\sigma)(e^{\hat{B}} )^N_{\ K} (\bar{\sigma}) \{\Pi^J (\sigma), \Pi^K(\bar{\sigma}) \} \simeq \kappa \, \Big[ e^B (\sigma) \eta e^{B^T} ({\bar \sigma}) \Big]^{M N} \delta^\prime (\sigma - {\bar \sigma} )    \, .
\end{eqnarray}  
Using the properties of the $\delta$ function
\begin{equation} \label{eq:fdelta}
f(\bar{\sigma}) \partial_\sigma \delta(\sigma-\bar{\sigma}) = f(\sigma) \partial_\sigma \delta(\sigma-\bar{\sigma})+f^\prime (\sigma) \delta(\sigma-\bar{\sigma}) \, ,
\end{equation} 
the term becomes
\begin{eqnarray} \label{eq:PBLan}
\kappa \eta^{MN}   \delta^{\prime}(\sigma-\bar{\sigma}) +\kappa (e^{\hat{B}})^M_{\ P} \eta^{PR} \partial_Q ((e^{\hat{B}})^T)_{R}^{\ N}  X^{\prime Q}\delta(\sigma-\bar{\sigma})\, ,
\end{eqnarray}
where we have used the orthogonal property of $B$-transformation (\ref{eq:eOe})
\begin{equation}
\kappa (e^{\hat{B}})^M_{\ P} \eta^{PR} ((e^{\hat{B}})^T)_{R}^{\ N} \delta^{\prime}(\sigma-\bar{\sigma})  = \kappa \eta^{MN}  \delta^{\prime}(\sigma-\bar{\sigma}) \, .
\end{equation}
After another application of T-duality, the other term becomes
\begin{equation} \label{eq:PBLan1}
\kappa (e^{\hat{B}})^M_{\ P} \eta^{PR} \partial_Q ((e^{\hat{B}})^T)_{R}^{\ N}  X^{\prime Q}\delta(\sigma-\bar{\sigma}) \simeq  (e^{\hat{B}})^M_{\ P} \partial_Q \hat{B}^{PN} \Pi^Q \delta(\sigma-\bar{\sigma}) \, .
\end{equation}

Using the properties of  $\hat{B}^M_{\ N}$ matrix  (\ref{eq:bhat})
\begin{eqnarray}\label{eq:Bprop}
\hat{B}^M{}_K \hat{B}^K{}_N = 0 \, , \qquad   \hat{B}^M{}_K  \partial^Q \hat{B}^K{}_N = 0 \, , \qquad  (e^{\hat{B}})^M{}_N = \delta^M{}_N + \hat{B}^M{}_N \, , \qquad
\end{eqnarray}
and substituting (\ref{eq:PiHat}) and (\ref{eq:PBLan}) into (\ref{eq:hphp1}), one obtains
\begin{eqnarray}\label{eq:PBL6}
\{ \hat{\Pi}^M (\sigma), \hat{\Pi}^N ({\bar \sigma}) \} = 
-   \hat{B}^{M N Q}   \, \hat{\Pi}_Q \delta (\sigma - {\bar \sigma} ) + A^{MN}(\sigma-\bar{\sigma})\, ,
\end{eqnarray}
where $ A^{MN}$ is the anomalous term, where
\begin{equation} \label{eq:anom-dual}
 A^{MN}(\sigma-\bar{\sigma}) \simeq \kappa \eta^{MN}\delta^{\prime}(\sigma-\bar{\sigma})    \, ,
\end{equation}
and $ \hat{B}^{MNQ}$  is the flux, where
\begin{eqnarray} \label{eq:BBS}
\hat{B}^{MNQ} &=& B^{MNQ} + S^{MNQ} \\ \notag
B^{M N Q}&=& \partial^M  \hat{B}^{N Q}  +    \partial^N  \hat{B}^{Q M} +    \partial^Q  \hat{B}^{M N} \\ \notag
S^{MNQ} &=&  \hat{B}^M{}_K   \partial^K  \hat{B}^{N Q}  + \hat{B}^N{}_K   \partial^K \hat{B}^{Q M} +\hat{B}^Q{}_K   \partial^K  \hat{B}^{M N} \, .
\end{eqnarray}
The flux can be also written in a more compact way as
\begin{eqnarray} \label{eq:BFlux}
 \hat{B}^{MNQ} = \hat{\partial}^M \hat{B}^{NQ} +  \hat{\partial}^N \hat{B}^{QM} + \hat{\partial}^Q \hat{B}^{MN} \, ,
\end{eqnarray}
where we have introduced new derivatives
\begin{eqnarray}\label{eq:hpartial}
\hat{\partial}^M  = (e^{\hat{B}})^M{}_K   \partial^K = \partial^M + \hat{B}^M_{\ K}\ \partial^K  \, .
\end{eqnarray}
The other necessary algebra is straightforwardly obtained 
\begin{eqnarray}\label{eq:PBL22}
\{ \hat{\Lambda}^M (\sigma), \hat{\Pi}^N ({\bar \sigma}) \} = \hat{\partial}^N   \hat{\Lambda}^M   \delta (\sigma - {\bar \sigma} ) \, .
\end{eqnarray}

The careful reader might notice that we wrote the equality instead of T-duality sign in the relations (\ref{eq:PBL6}), even though the relations between the momenta $\Pi^M$ were given in terms of their T-duals (\ref{eq:PBPX}). This is possible since the two successive applications of T-duality act as the identity operation. However, on the anomalous part we actually applied T-duality relations only once, the second application will follow soon.

Now we are ready to calculate the full bracket
\begin{eqnarray}\label{eq:DefC1}
\{ {\cal \hat{G}}_{\hat{\Lambda}_1} (\sigma), {\cal \hat{G}}_{\hat{\Lambda}_2} ({\bar \sigma}) \} &=&
\hat{\Lambda}_1^M   (\sigma)  \hat{\Lambda}_2^N  ({\bar \sigma}) A_{MN}- \hat{\Lambda}_{1 M}  \hat{\Lambda}_{2 N}   \hat{B}^{M N Q}    \hat{\Pi}_Q \delta (\sigma - {\bar \sigma} ) 
\\ \notag
&&+ \hat{\Pi}_Q   \Big[ \hat{\Lambda}_2^N  \hat{\partial}_N  \hat{\Lambda}_1^Q   -
\hat{\Lambda}_1^N  \hat{\partial}_N \hat{\Lambda}_2^Q  \Big]  \delta (\sigma - {\bar \sigma} )  \, .
\end{eqnarray}
Again, using (\ref{eq:fdelta}) and (\ref{eq:anom-dual}), we obtain
\begin{eqnarray}\label{eq:DefC2}
&&\hat{\Lambda}_1^M   (\sigma)  \hat{\Lambda}_{2}^N (\bar{\sigma}) A_{MN} (\sigma-\bar{\sigma})  \\ \notag
&&\simeq \kappa
 \langle \hat{\Lambda}_1   (\sigma) , \hat{\Lambda}_{2}  ({\sigma})\rangle  \delta^\prime (\sigma - {\bar \sigma} ) + \kappa\langle
\hat{\Lambda}_1   (\sigma)  , \hat{\Lambda}_{2}^{\prime}  ({\sigma}) \rangle  \delta (\sigma - {\bar \sigma} )  \\ \notag
&&= \frac{\kappa}{2} \Big( 2\langle \hat{\Lambda}_1  ,  \hat{\Lambda}_2 \rangle   \delta^\prime (\sigma - {\bar \sigma} ) +  \langle \hat{\Lambda}_1 , \hat{\Lambda}_2 \rangle^\prime   \delta (\sigma - {\bar \sigma} )\Big)
+ \frac{\kappa}{2} \Big( \langle \hat{\Lambda}_1  ,  \hat{\Lambda}_2^{\prime}\rangle  -\langle \hat{\Lambda}_1^{\prime},    \hat{\Lambda}_2 \rangle \Big) \delta (\sigma - {\bar \sigma} ) \\ \notag
&&= \frac{\kappa}{2} \Big( \langle \hat{\Lambda}_1  ,  \hat{\Lambda}_2 \rangle (\sigma)  +  \langle \hat{\Lambda}_1  ,  \hat{\Lambda}_2 \rangle (\bar{\sigma})  \Big)  \delta^\prime (\sigma - {\bar \sigma} ) 
+ \frac{\kappa}{2} \Big( \langle \hat{\Lambda}_1  ,  \hat{\Lambda}_2^{\prime}\rangle  -\langle \hat{\Lambda}_1^{\prime},    \hat{\Lambda}_2 \rangle \Big) \delta (\sigma - {\bar \sigma} ) \, .
\end{eqnarray}
The anomalous part correspond exactly to the anomalous part of the relation (\ref{eq:algdef}). It does not contribute to the twisted $C$-bracket. Using the T-duality relations (\ref{eq:TD1}) and definitions (\ref{eq:skalproizvod})  and (\ref{eq:hpartial}), we write the remaining part as
\begin{eqnarray}\label{eq:LL}
\frac{\kappa}{2}  \Big( \langle \hat{\Lambda}_1, \hat{\Lambda}_2^\prime    \rangle  -  \langle \hat{\Lambda}_1^\prime , \hat{\Lambda}_2   \rangle \Big) &=&  \frac{\kappa}{2}  \eta_{M N} \,  \Big( \hat{\Lambda}_1^M \partial_Q \hat{\Lambda}_2^N    - \hat{\Lambda}_2^M \partial_Q \hat{\Lambda}_1^N  \Big)  X^{\prime Q}  \\ \notag
  &\simeq& \frac{1}{2}  \eta_{M N} \,  \Big( \hat{\Lambda}_1^M \partial^Q \hat{\Lambda}_2^N    - \hat{\Lambda}_2^M \partial^Q \hat{\Lambda}_1^N  \Big)  \Pi_Q \\ \notag
&=& \frac{1}{2}  \eta_{M N} \,  \Big( \hat{\Lambda}_1^M \hat{\partial}^Q \hat{\Lambda}_2^N    - \hat{\Lambda}_2^M \hat{\partial}^Q \hat{\Lambda}_1^N  \Big)
\hat{\Pi}_Q  \, .
\end{eqnarray} 
Therefore, we applied the T-duality the second time on the last term contributing to the $B$-twisted $C$-bracket. The full bracket is given by
\begin{eqnarray}\label{eq:C-B}
{[\hat{\Lambda}_1, \hat{\Lambda}_2]_{{\textbf C}_B}}^Q &=&
\hat{\Lambda}_1^N     \hat{\partial}_N  \hat{\Lambda}_2^Q
- \hat{\Lambda}_2^N     \hat{\partial}_N   \hat{\Lambda}_1^Q
  \\ \notag
&& - \frac{1}{2}  \eta_{M N} \,  \Big( \hat{\Lambda}_1^M \hat{\partial}^Q \hat{\Lambda}_2^N    - \hat{\Lambda}_2^M \hat{\partial}^Q \hat{\Lambda}_1^N   \Big)
 + \hat{\Lambda}_{1 M}  \hat{\Lambda}_{2 N}   \hat{B}^{M N Q}  \, .
\end{eqnarray}
This is the $C$-bracket twisted by $B$. Substituting the expansion of derivative (\ref{eq:hpartial}), we can separate terms that contain $B$ from those that do not 
\begin{eqnarray}
{[\hat{\Lambda}_1, \hat{\Lambda}_2]_{{\textbf C}_B}}^Q &=& \hat{\Lambda}_1^N     \partial_N  \hat{\Lambda}_2^Q- \hat{\Lambda}_2^N     \partial_N   \hat{\Lambda}_1^Q - \frac{1}{2}  \eta_{M N} \,  \Big( \hat{\Lambda}_1^M \partial^Q \hat{\Lambda}_2^N    - \hat{\Lambda}_2^M \partial^Q \hat{\Lambda}_1^N   \Big) \\ \notag
&& +\hat{B}^N_{\ R}\Big(\hat{\Lambda}_{1N} \partial^R \hat{\Lambda}_{2}^Q-\hat{\Lambda}_{2N} \partial^R \hat{\Lambda}_{1}^Q \Big) - \frac{1}{2}\hat{B}^Q_{\ R}\Big(\hat{\Lambda}_{1N} \partial^R \hat{\Lambda}_{2}^N-\hat{\Lambda}_{2N} \partial^R \hat{\Lambda}_{1}^N \Big) \\ \notag
&&  + \hat{\Lambda}_{1 M}  \hat{\Lambda}_{2 N}   \hat{B}^{M N Q} 
\end{eqnarray}
The first line is the $C$-bracket, while the other two lines are corrections due to twisting. If the Kalb-Ramond field is zero, the bracket reduces to the $C$-bracket. 

In \cite{GSM}, authors discussed the conditions under which $B$-shifts are automorphisms of the $C$-bracket, i.e. $e^{\hat{B}}[\Lambda_1, \Lambda_2]_{\bf C} = [e^{\hat{B}} \Lambda_1, e^{\hat{B}} \Lambda_2]_{\bf C}$, which they found to be correct for $\hat{B}^M_{\ N} \partial^N= 0$ and $\partial_M \hat{B}_{NR} + cyclic = 0$. We considered a more general case of arbitrary $B$-shifts and obtained $B$-twisted $C$-bracket, from which above conditions for automorphism can be read directly.

\section{C-bracket twisted by $\theta$}
\cleq

In this chapter, we will show how the $\theta$-twisted $C$-bracket can be obtained from the generator's algebra. By analogy, consider the string moving in the background characterized only with the T-dual metric tensor. The generalized metric is given by
\begin{equation} \label{eq:dualGMN}
{}^{\star} G_{MN} = \begin{pmatrix}
{}^\star G^{-1}_{\mu \nu} & 0 \\
0 & {}^\star G^{\mu \nu}
\end{pmatrix} = 
\begin{pmatrix}
G^E_{\mu \nu} & 0 \\
0 & (G_E^{-1})^{\mu \nu}
\end{pmatrix} \, ,
\end{equation}
where $G_E$ is defined in (\ref{eq:Geff}). Now let us consider another important $O(D,D)$ transformation realized by 
\begin{equation} \label{eq:enateta}
(e^{\hat{\theta}})^M_{\ N} = 
\begin{pmatrix}
 \delta^\mu_\nu & \kappa \theta^{\mu \nu} \\
0 & \delta^\nu_\mu
\end{pmatrix} \, ,
\end{equation}
where
\begin{equation}
\hat{ \theta}^M_{\ N} = 
\begin{pmatrix}
0 & \kappa \theta^{\mu \nu} \\
0 & 0 
\end{pmatrix} = 
\begin{pmatrix}
0 & 2 \ {^\star B}^{\mu \nu} \\
0 & 0
\end{pmatrix} \, .
\end{equation}
This is known as a $\theta$-transformation. We have
\begin{equation} \label{eq:eminteta}
 ( (e^{\hat{\theta}})^T )_{M}^{\ N} =
\begin{pmatrix}
\delta^\nu_\mu & 0\\
-\kappa \theta^{\mu \nu}  & \delta^\mu_\nu  
\end{pmatrix} \, ,
\end{equation}
from which we verify that it is indeed $O(D,D)$
\begin{equation} \label{eq:etOet}
( (e^{\hat{\theta}})^T )_{M}^{\ L}\ \eta_{LK}\ (e^{\hat{\theta}})^K_{\ N} = \eta_{MN} \, .
\end{equation}
Under this transformation, the diagonal generalized metric (\ref{eq:dualGMN}) goes to 
\begin{equation}
{}^\star H_{MN} = ( (e^{\hat{\theta}})^T )_{M}^{\ L}\ {^\star G}_{LK} (e^{\hat{\theta}})^K_{\ N} = 
\begin{pmatrix}
G^E_{\mu \nu}& - 2B_{\mu \rho} (G^{-1})^{\rho \nu} \\
 2(G^{-1})^{\mu \rho} B_{\rho \nu} & (G^{-1})^{\mu \nu}
\end{pmatrix} \, ,
\end{equation}
which is exactly equal to the generalized metric (\ref{eq:genmet}).

The Hamiltonian (\ref{eq:Hcan}) can be rewritten in terms of ${}^{\star} G_{MN}$
\begin{equation}
{\cal H}_{\cal C} = \frac{1}{2\kappa} \breve{\Pi}_M  {}^{\star} G^{MN} \breve{\Pi}_N + \frac{\kappa}{2} \breve{X}^{\prime M}{}^{\star}  G_{MN} \breve{X}^{\prime N} \, ,
\end{equation}
where 
\begin{equation}
 \breve{X}^{\prime M}  = (e^{\hat{\theta}})^M{}_N  X^{\prime N}   =
\begin{pmatrix}
 \delta^\mu_\nu & \kappa \theta^{\mu \nu} \\
0 & \delta^\nu_\mu
\end{pmatrix} \, 
\begin{pmatrix}
x^{\prime \nu}     \\
y^{\prime}_\nu   \\
\end{pmatrix} = 
\begin{pmatrix}
x^{\prime \mu} + \kappa \theta^{\mu \nu} y_{\nu}^{\prime} \\
y_{\mu}^{\prime}
\end{pmatrix}\equiv 
\begin{pmatrix}
\breve{x}^{\prime \mu} \\
y^{\prime}_{\mu}
\end{pmatrix}\, .
\end{equation}
and
\begin{eqnarray}\label{eq:Pibreve}
 \breve{\Pi}^M  = (e^{\hat{\theta}})^M{}_N  \Pi^N   =
\begin{pmatrix}
 \delta^\mu_\nu & \kappa \theta^{\mu \nu} \\
0 & \delta^\nu_\mu
\end{pmatrix} \, 
\begin{pmatrix}
{}^\star \pi^\nu     \\
\pi_\nu   \\
\end{pmatrix}
=
\begin{pmatrix}
{}^\star \pi^\mu + \kappa \theta^{\mu \nu} \pi_\nu    \\
\pi_\mu  \\
\end{pmatrix}
\equiv
\begin{pmatrix}
{}^\star \breve{\pi}^\mu     \\
\pi_\mu    \\
\end{pmatrix}
   \, .
\end{eqnarray}
The new symmetry parameters are given by
\begin{eqnarray}\label{eq:Lbreve}
 \breve{\Lambda}^M  = (e^{\hat{\theta}})^M{}_N\Lambda^N  =
\begin{pmatrix}
 \delta^\mu_\nu & \kappa \theta^{\mu \nu} \\
0 & \delta^\nu_\mu
\end{pmatrix} \, 
\begin{pmatrix}
   \xi^\nu     \\
\lambda_\nu   \\
\end{pmatrix}
=
\begin{pmatrix}
 \xi^\mu + \kappa \theta^{\mu \nu} \lambda_\nu    \\
\lambda_\mu  \\
\end{pmatrix}
\equiv
\begin{pmatrix}
\breve{\xi}^\mu     \\
 \lambda_\mu   \\
\end{pmatrix} \, ,
\end{eqnarray}
in terms of which we write the symmetry generator as
\begin{equation}
\breve{\cal G}_{\breve{\Lambda}}  =\breve{\Lambda}^M \eta_{MN} \breve{\Pi}^N \, .
\end{equation}
Its algebra results in the $\theta$-twisted $C$-bracket
\begin{equation} \label{eq:basssic}
\{ \breve{\cal G}_{\breve{\Lambda}_1} (\sigma), \breve{\cal G}_{\breve{\Lambda}_2} (\bar{\sigma}) \} =  - \breve{\cal G}_{[ \breve{\Lambda}_1,  \breve{\Lambda}_2]_{{\textbf C}_\theta}}(\sigma) \delta(\sigma-\bar{\sigma})   +\frac{\kappa}{2} \Big(\langle \breve{\Lambda}_1, \breve{\Lambda}_2 \rangle (\sigma) + \langle \breve{\Lambda}_1, \breve{\Lambda}_2 \rangle (\bar{\sigma}) \Big) \delta^{\prime}(\sigma-\bar{\sigma}) \, ,
\end{equation}
where 
\begin{equation} \label{eq:}
[ \breve{\Lambda}_1,  \breve{\Lambda}_2]_{{\textbf C}_\theta} =  e^{\hat{\theta}}  [e^{-\hat{\theta}} \breve{\Lambda}_1, e^{-\hat{\theta}}
 \breve{\Lambda}_2]_{{\textbf C}}  \, .
\end{equation}

Let us outline the algebra relations necessary for obtaining this bracket. Firstly, we have
\begin{eqnarray} \label{eq:PPB1}
\{\breve{\Pi}^M (\sigma), \breve{\Pi}^N  (\bar{\sigma})\} &=& \{ (e^{\hat{\theta}}  \Pi )^M (\sigma), (e^{\hat{\theta}}  \Pi )^N  (\bar{\sigma}) \} \\ \notag
&=& (e^{\hat{\theta}} )^M_{\ J} (\sigma)(e^{\hat{\theta}} )^N_{\ K} (\bar{\sigma}) \{\Pi^J (\sigma), \Pi^K(\bar{\sigma}) \} \\ \notag
&&- (e^{\hat{\theta}})^M_{\ J}\partial^J (e^{\hat{\theta}})^N_{\ Q} \Pi^Q \delta(\sigma-\bar{\sigma}) + (e^{\hat{\theta}})^N_{\ J} \partial^J (e^{\hat{\theta}})^M_{\ Q} \Pi^Q \delta(\sigma-\bar{\sigma}) \, .
\end{eqnarray}
Next, using (\ref{eq:PBPX}) and (\ref{eq:fdelta}), we obtain
\begin{eqnarray} \label{eq:PPB2}
(e^{\hat{\theta}} )^M_{\ J} (\sigma)(e^{\hat{\theta}} )^N_{\ K} (\bar{\sigma}) \{\Pi^J (\sigma), \Pi^K(\bar{\sigma}) \}  =  A^{MN}(\sigma-\bar{\sigma}) + (e^{\hat{\theta}})^M_{\ P} \partial_Q \hat{\theta}^{PN} \Pi^Q \delta(\sigma-\bar{\sigma}) \, ,
\end{eqnarray}
where $A^{MN}$ is the same anomaly as in the previous chapter (\ref{eq:anom-dual}). Substituting (\ref{eq:PPB2}) and (\ref{eq:Pibreve}) into (\ref{eq:PPB1}), we obtain
\begin{equation} \label{eq:bassic2}
\{\breve{\Pi}^M (\sigma), \breve{\Pi}^N  (\bar{\sigma})\} = - \breve{\Theta}^{MNQ}\breve{\Pi}_Q \delta(\sigma-\bar{\sigma}) + A^{MN}(\sigma-\bar{\sigma})\, , 
\end{equation}
where 
\begin{eqnarray} \label{eq:TTR}
\breve{\Theta}^{MNQ} &=& {\Theta}^{MNQ} + R^{MNQ} \\ \notag
{\Theta}^{M N Q}&=& \partial^M  \hat{\theta}^{N Q}  +    \partial^N  \hat{\theta}^{Q M} +    \partial^Q  \hat{\theta}^{M N} \\ \notag
R^{MNQ} &=&  \hat{\theta}^M{}_K   \partial^K  \hat{\theta}^{N Q}  + \hat{\theta}^N{}_K   \partial^K \hat{\theta}^{Q M} +\hat{\theta}^Q{}_K   \partial^K  \hat{\theta}^{M N} \, .
\end{eqnarray}
This flux can be also written as
\begin{equation} \label{eq:breveTh}
\breve{\Theta}^{MNR} =  \breve{\partial}^M \hat{\theta}^{NR} + \breve{\partial}^N \hat{\theta}^{RM} +\breve{\partial}^R \hat{\theta}^{MN} \, .
\end{equation}
where $\breve{\partial}^M$ are new derivatives, given by
\begin{equation} \label{eq:brevepartial}
\breve{\partial}^M = (e^{\hat{\theta}})^M_{\ N}\ \partial^N = \partial^M + \hat{\theta}^M_{\ N}\ \partial^N  \, .
\end{equation}
Next, we have
\begin{equation} \label{eq:bassic3}
\{  \breve{\Lambda}^M (\sigma), \breve{\Pi}^N (\bar{\sigma}) \} = \breve{\partial}^N \breve{\Lambda}^M \delta(\sigma-\bar{\sigma}) \, .
\end{equation}

Comparing relations (\ref{eq:bassic2}) to (\ref{eq:PBL6}), and (\ref{eq:bassic3}) to (\ref{eq:PBL22}), we see that the basic algebra relations have the same form. Therefore, we write
\begin{eqnarray}\label{eq:1DefC} \notag
\{ {\cal \breve{G}}_{\breve{\Lambda}_1} (\sigma), {\cal \breve{G}}_{\breve{\Lambda}_2} ({\bar \sigma}) \} &=&
\breve{\Lambda}_1^M   (\sigma)  \breve{\Lambda}_2^N  ({\bar \sigma}) A_{MN}(\sigma-\bar{\sigma})- \breve{\Lambda}_{1 M}  \breve{\Lambda}_{2 N}   \breve{\Theta}^{M N Q}    \breve{\Pi}_Q \delta (\sigma - {\bar \sigma} ) \\ 
&&+ \breve{\Pi}_Q   \Big[ \breve{\Lambda}_2^N  \breve{\partial}_N  \breve{\Lambda}_1^Q   -
\breve{\Lambda}_1^N  \breve{\partial}_N \breve{\Lambda}_2^Q  \Big]  \delta (\sigma - {\bar \sigma} )  \, ,
\end{eqnarray}
and after using (\ref{eq:DefC2}) and (\ref{eq:LL}), we obtain the full $\theta$-twisted $C$-bracket
\begin{eqnarray}\label{eq:C-Th}
{[\breve{\Lambda}_1, \breve{\Lambda}_2]_{{\textbf C}_\theta}}^Q &=&
\breve{\Lambda}_1^N     \breve{\partial}_N  \breve{\Lambda}_2^Q
- \breve{\Lambda}_2^N     \breve{\partial}_N   \breve{\Lambda}_1^Q
  \\ \notag
&& - \frac{1}{2}  \eta_{M N} \,  \Big( \breve{\Lambda}_1^M \breve{\partial}^Q \breve{\Lambda}_2^N    - \breve{\Lambda}_2^M \breve{\partial}^Q \breve{\Lambda}_1^N   \Big)
 + \breve{\Lambda}_{1 M}  \breve{\Lambda}_{2 N}   \breve{\Theta}^{M N Q}  \, .
\end{eqnarray}
The only difference between the $B$- and $\theta$-twisted $C$-brackets is that the derivatives are defined with $B$-shifts in the former and $\theta$-transformations in the latter case, and the flux term is given in terms of different, mutually T-dual fields. Due to the T-duality relations between these fields (\ref{eq:GBstar}), the two twisted $C$-brackets are mutually T-dual as well.

Using (\ref{eq:brevepartial}), we can rewrite the above bracket as
\begin{eqnarray}
{[\breve{\Lambda}_1, \breve{\Lambda}_2]_{{\textbf C}_{\theta}}}^Q &=& \breve{\Lambda}_1^N     \partial_N  \breve{\Lambda}_2^Q- \breve{\Lambda}_2^N     \partial_N   \breve{\Lambda}_1^Q - \frac{1}{2}  \eta_{M N} \,  \Big( \breve{\Lambda}_1^M \partial^Q \breve{\Lambda}_2^N    - \breve{\Lambda}_2^M \partial^Q \breve{\Lambda}_1^N   \Big) \\ \notag
&& +\hat{\theta}^N_{\ R}\Big(\breve{\Lambda}_{1N} \partial^R \breve{\Lambda}_{2}^Q-\breve{\Lambda}_{2N} \partial^R \breve{\Lambda}_{1}^Q \Big) - \frac{1}{2}\hat{\theta}^Q_{\ R}\Big(\breve{\Lambda}_{1N} \partial^R \breve{\Lambda}_{2}^N-\breve{\Lambda}_{2N} \partial^R \breve{\Lambda}_{1}^N \Big) \\ \notag
&&  + \breve{\Lambda}_{1 M}  \breve{\Lambda}_{2 N}   \breve{\Theta}^{M N Q}  
\end{eqnarray}
The first line is the $C$-bracket, while the remaining terms are corrections due to its twisting. If the bi-vector $\theta$ is zero, this bracket reduces to the standard $C$-bracket.

From the double theory it is possible to obtain both the initial $\sigma$ model, as well its T-dual description. The former is obtained when all fields and parameters depend on the coordinate $x^\mu$ only, and the latter when they depend only on $y_\mu$. In the following chapters, we analyze the twisted $C$-brackets realization in these theories.

\section{Initial theory}
\cleq 

Firstly consider the $B$-twisted $C$-bracket (\ref{eq:C-B}) and its projection to the initial theory, characterized solely by coordinates $x^\mu$. If the parameters $\hat{\Lambda}$ and Kalb-Ramond field $B$ do not depend on the T-dual coordinate $y$, the derivatives $\hat{\partial}^Q$ become just derivatives along the initial coordinates $x^\mu$
\begin{equation}
\hat{\partial}^Q \to \begin{pmatrix}
\delta^\mu_\nu & 0 \\
2 B_{\mu \nu} & \delta^\nu_\mu 
\end{pmatrix} 
\begin{pmatrix}
0 \\
\partial_\nu 
\end{pmatrix} =
\begin{pmatrix}
0 \\
\partial_\mu 
\end{pmatrix} \, .
\end{equation}
It is easy to obtain that
\begin{equation} \label{eq:cb1}
\hat{\Lambda}_1^N     \hat{\partial}_N  \hat{\Lambda}_2^Q \to 
\begin{pmatrix}
\xi_1^\nu \partial_\nu \xi_2^\mu \\ 
 \xi_1^\nu \partial_\nu \hat{\lambda}_{2\mu} 
\end{pmatrix}\, ,
\end{equation}
and 
\begin{equation}\label{eq:cb2}
\eta_{M N} \hat{\Lambda}_1^M \hat{\partial}^Q \hat{\Lambda}_2^N \to 
\begin{pmatrix}
0 \\
\hat{\lambda}_{1\nu} \partial_\mu \xi_2^\nu + \xi_1^\nu \partial_\mu \hat{\lambda}_{2\nu}
\end{pmatrix} \, .
\end{equation}

We see that the derivative $\hat{\partial}^Q$ no longer depends on the Kalb-Ramond field. The flux $\hat{B}^{MNQ}$ therefore reduces to $B^{MNQ}$, and
\begin{equation}
\hat{B}^{MNQ} \hat{\Lambda}_{1M} \hat{\Lambda}_{2N} \to \begin{pmatrix}
0 \\
2 B_{\mu \nu \rho} \xi^{\nu}_1 \xi^{\rho}_2
\end{pmatrix} \, ,
\end{equation}
where $B_{\mu \nu \rho}$ is the Kalb-Ramond field strength, given by
\begin{equation}
B_{\mu \nu \rho} = \partial_\mu B_{\nu \rho} +\partial_\nu B_{\rho \mu} +\partial_\rho B_{\mu \nu} \, .
\end{equation}
This term represents the geometric $H$ flux in the initial theory \cite{flux1}.  
Combining previous relations and using the chain rule, the $B$-twisted $C$-bracket becomes 
\begin{equation}
[\hat{\Lambda}_1, \hat{\Lambda}_2]_{{\textbf C}_B} \to [\hat{\Lambda}_1, \hat{\Lambda}_2]_{{\cal C}_B} = \hat{\Lambda} \equiv \begin{pmatrix}
\xi \\
\hat{\lambda}
\end{pmatrix}\, ,
\end{equation}
where
\begin{eqnarray} \label{eq:xiLB}
\xi^\mu &=& \xi_1^\nu \partial_\nu \xi_2^\mu - \xi_2^\nu \partial_\nu \xi_1^\mu \, ,\\ \notag 
\hat{\lambda}_\mu &=&\xi_1^\nu (\partial_\nu \hat{\lambda}_{2 \mu} - \partial_\mu \hat{\lambda}_{2 \nu}) - \xi_2^\nu (\partial_\nu \hat{\lambda}_{1 \mu} - \partial_\mu \hat{\lambda}_{1 \nu}) +\frac{1}{2} \partial_\mu (\xi_1 \hat{\lambda}_2- \xi_2 \hat{\lambda}_1 )+ 2  B_{\mu \nu \rho} \xi^\nu_1 \xi^\rho_2 \, .
\end{eqnarray} 
This bracket is a well known Courant bracket twisted by a 2-form $B$ \cite{twist}.

Now let us consider the realization of the $\theta$-twisted $C$-bracket (\ref{eq:C-Th}) in the initial phase space. If we omit the dependence form $y_\mu$, the derivative $\breve{\partial}$ becomes
\begin{equation}
\breve{\partial}^Q \to \begin{pmatrix}
\delta^\mu_\nu & \kappa \theta^{\mu \nu} \\
0 & \delta^\nu_\mu 
\end{pmatrix} 
\begin{pmatrix}
0 \\
\partial_\nu 
\end{pmatrix} =
\begin{pmatrix}
\kappa \theta^{\mu \nu}\partial_\nu  \\
\partial_\mu 
\end{pmatrix} \, .
\end{equation} 
Now, we have
\begin{equation} \label{eq:cth1}
\breve{\Lambda}_1^N     \breve{\partial}_N  \breve{\Lambda}_2^Q \to 
\begin{pmatrix} 
\lambda_{1\nu} \kappa \theta^{\nu \rho } \partial_\rho   \breve{\xi}_2^\mu + \breve{\xi}_1^\nu \partial_\nu \breve{\xi}_2^\mu \\ 
\lambda_{1\nu} \kappa \theta^{\nu \rho } \partial_{\rho} \lambda_{2\mu} + \breve{\xi}_1^{\nu} \partial_{\nu} \lambda_{2\mu}
\end{pmatrix}\, ,
\end{equation}
and 
\begin{equation}\label{eq:cth2}
\eta_{M N} \breve{\Lambda}_1^M \breve{\partial}^Q \breve{\Lambda}_2^N \to 
\begin{pmatrix}
\kappa \theta^{\mu \nu}(\breve{\xi}_1^\rho \partial_\nu \lambda_{2\rho} + \lambda_{1\rho} \partial_\nu \breve{\xi}^\rho_2) \\
\breve{\xi}_1^\rho \partial_\mu \lambda_{2\rho} + \lambda_{1\rho} \partial_\mu \breve{\xi}^\rho_2 
\end{pmatrix} \, .
\end{equation}
The flux term is given by
\begin{equation} \label{eq:cth3}
\breve{\Lambda}_{1 M}  \breve{\Lambda}_{2 N}   \breve{\Theta}^{M N Q} \to 
\begin{pmatrix}
\kappa^2 R^{\mu \nu \rho} \lambda_{1 \nu}\lambda_{2 \rho} +\kappa Q_\nu^{\ \rho \mu} ( \breve{\xi}_1^\nu \lambda_{2\rho}-\breve{\xi}_2^\nu \lambda_{1\rho}) \\
\kappa Q_\mu^{\ \rho \nu} \lambda_{1\rho} \lambda_{2\nu}  
\end{pmatrix}\, ,
\end{equation}
where $Q$ and $R$ are non-geometric fluxes \cite{flux1}, given by
\begin{equation} \label{eq:QRdef}
Q_\mu^{\ \nu \rho} = \partial_\mu \theta^{\nu \rho} \, , \quad R^{\mu \nu \rho} = \theta^{\mu \sigma} \partial_\sigma \theta^{\nu \rho}+\theta^{\nu \sigma} \partial_\sigma \theta^{\rho \mu}+\theta^{\rho \sigma} \partial_\sigma \theta^{\mu \nu} \, .
\end{equation}
We see that $\Theta^{MNQ}$ gave rise to non-geometric $Q$-flux, while $R^{MNQ}$ gave rise to the non-geometric $R$-flux. 

Substituting (\ref{eq:cth1}), (\ref{eq:cth2}) and (\ref{eq:cth3}) into (\ref{eq:C-Th}), with the help of chain rule, we obtain
\begin{equation}
[ \breve{\Lambda}_1, \breve{\Lambda}_2 ]_{{\textbf C}_{\theta}} \to[ \breve{\Lambda}_1, \breve{\Lambda}_2 ]_{\cal{C}_{\theta}}  = \breve{\Lambda} \equiv \begin{pmatrix}
\breve{\xi} \\
\lambda
\end{pmatrix} \, ,
\end{equation}
where
\begin{eqnarray} \label{eq:xiLR}
\breve{\xi}^\mu &=&\ \breve{\xi}_1^\nu \partial_\nu \breve{\xi}_2^\mu - \breve{\xi}_2^\nu \partial_\nu\breve{\xi}_1^\mu + \\ \notag
&& -\kappa \theta^{\mu \nu}\Big( \breve{\xi}_1^\rho (\partial_\nu \lambda_{2 \rho}-\partial_\rho \lambda_{2 \nu}) - \breve{\xi}_2^\rho ( \partial_\nu \lambda_{1 \rho}-\partial_\rho \lambda_{1 \nu}) -\frac{1}{2} \partial_\nu (\breve{\xi}_1 \lambda_{2} - \breve{\xi}_2 \lambda_1) \Big) \\ \notag
&& + \kappa \breve{\xi}_1^\nu \partial_\nu (\lambda_{2 \rho} \theta^{\rho \mu})-\kappa \breve{\xi}_2^\nu \partial_\nu (\lambda_{1 \rho} \theta^{\rho \mu})+\kappa (\lambda_{1 \nu} \theta^{\nu \rho}) \partial_\rho \breve{\xi}_2^\mu -\kappa (\lambda_{2 \nu}\theta^{\nu \rho}) \partial_\rho \breve{\xi}_1^\mu \\ \notag
&&+\kappa^2 (\theta^{\mu \sigma} \partial_\sigma \theta^{\nu \rho}+\theta^{\nu \sigma} \partial_\sigma \theta^{\rho \mu}+\theta^{\rho \sigma} \partial_\sigma \theta^{\mu \nu}) \lambda_{1 \nu}\lambda_{2 \rho} \, , \\ \notag 
\lambda_\mu &= &\ \breve{\xi}_1^\nu (\partial_\nu \lambda_{2 \mu} - \partial_\mu \lambda_{2 \nu}) - \breve{\xi}_2^\nu (\partial_\nu \lambda_{1 \mu} - \partial_\mu \lambda_{1 \nu}) +\frac{1}{2}\partial_\mu(\breve{\xi}_1 \lambda_2 - \breve{\xi}_2 \lambda_1) \\ \notag
&& + \kappa \theta^{\nu \rho} (\lambda_{1 \nu}\partial_\rho \lambda_{2 \mu}-\lambda_{2 \nu} \partial_\rho \lambda_{1 \mu})+ \kappa \lambda_{1 \rho} \lambda_{2 \nu} \partial_\mu \theta^{\rho \nu} \, .
\end{eqnarray}
This is the Courant bracket twisted by a bi-vector $\theta$.  As expected, the twisted $C$-brackets reduce to their twisted Courant counterparts in the initial theory, in the same way that the $C$-bracket reduces to the Courant one, once there is no dependence on $y_\mu$. 

\section{T-dual theory}
\cleq

Now let us consider the twisted $C$-brackets in the T-dual theory, that is to say the theory obtained after applying the T-dualization along all of the initial coordinates $x^\mu$. Effectively, this is obtained when we demand all quantities to depend solely on $y_\mu$. 

For the $B$-twisted $C$-bracket, we have
\begin{equation}
\hat{\partial}^Q \to \begin{pmatrix}
\delta^\mu_\nu & 0 \\
2 B_{\mu \nu} & \delta^\nu_\mu 
\end{pmatrix} 
\begin{pmatrix}
\tilde{\partial}^\nu \\
0
\end{pmatrix} =
\begin{pmatrix}
\tilde{\partial}^\mu \\
2 B_{\mu \nu} \tilde{\partial}^\nu
\end{pmatrix} \, .
\end{equation}
Moreover, we obtain
\begin{equation} \label{eq:cbd1}
\hat{\Lambda}_1^N     \hat{\partial}_N  \hat{\Lambda}_2^Q \to 
\begin{pmatrix}
\hat{\lambda}_{1 \nu} \tilde{\partial}^\nu \xi_2^\mu + 2 B_{\nu \rho} \xi^\rho_1 \tilde{\partial}^\nu \xi_2^\mu \\ 
 \hat{\lambda}_{1\nu} \tilde{\partial}^\nu \hat{\lambda}_{2\mu} + 2 B_{\nu \rho} \xi^\rho_1 \tilde{\partial}^\nu \hat{\lambda}_{2\mu}
\end{pmatrix}\, ,
\end{equation}
and 
\begin{equation}\label{eq:cbd2}
\eta_{M N} \hat{\Lambda}_1^M \hat{\partial}^Q \hat{\Lambda}_2^N \to 
\begin{pmatrix}
\hat{\lambda}_{1\nu} \tilde{\partial}^\mu \xi_2^\nu + \xi_1^\nu \tilde{\partial}^\mu \hat{\lambda}_{2\nu} \\
2\hat{\lambda}_{1\nu} B_{\mu \rho}\tilde{\partial}^\rho \xi_2^\nu +2 \xi_1^\nu B_{\mu \rho} \tilde{\partial}^\rho \hat{\lambda}_{2\nu}
\end{pmatrix} \, .
\end{equation}

The term containing flux $\hat{B}^{MNQ}$ becomes
\begin{equation}
\hat{B}^{MNQ} \hat{\Lambda}_{1M} \hat{\Lambda}_{2N} \to \begin{pmatrix}
\kappa\ {{}^\star Q}^\mu_{\ \nu \rho} \xi_1^\nu \xi_2^\rho  \\
 \kappa\ {{}^\star Q}^\nu_{\ \rho\mu}  (\xi_1^\rho \hat{\lambda}_{2\nu} -\xi_2^\rho \hat{\lambda}_{1\nu} ) +
\kappa^2 {{}^\star R}_{\mu \nu \rho} \xi^{\nu}_1 \xi^{\rho}_2
\end{pmatrix} \, ,
\end{equation}
where we have marked the non-geometric fluxes in T-dual theory by
\begin{equation}
\kappa\ {{}^\star Q}_{\ \nu \rho}^{\mu} = 2 \tilde{\partial}^\mu B_{\nu \rho} = \kappa\ \tilde{\partial}^\mu \  {{}^\star \theta}_{\nu \rho} \, ,
\end{equation}
and
\begin{eqnarray}
\kappa^2 {{}^\star R}_{\mu \nu \rho} &=&4 B_{\mu \sigma} \tilde{\partial}^\sigma B_{\nu \rho} +4B_{\nu \sigma} \tilde{\partial}^\sigma B_{\rho \mu}+4B_{\rho \sigma} \tilde{\partial}^\sigma B_{\mu \nu} \, , \\ \notag
&=& \kappa^2\ {{}^\star \theta}_{\mu \sigma} \tilde {\partial}^\sigma \ {{}^\star \theta}_{\nu \rho} +\kappa^2 \ {{}^\star \theta}_{\nu \sigma} \tilde{\partial}^\sigma \ {{}^\star \theta}_{\rho \mu}+\kappa^2 \ {{}^\star \theta}_{\rho \sigma} \tilde{\partial}^\sigma \ {{}^\star \theta}_{\mu \nu} \, .
\end{eqnarray}
Combining previous relations and using the chain rule, the $B$-twisted $C$-bracket becomes 
\begin{equation}
[\hat{\Lambda}_1, \hat{\Lambda}_2]_{{\textbf C}_B} \to [\hat{\Lambda}_1, \hat{\Lambda}_2]_{{\cal{C}}_B}  = \hat{\Lambda} \equiv \begin{pmatrix}
\xi \\
\hat{\lambda}
\end{pmatrix}\, ,
\end{equation}
where
\begin{eqnarray} \label{eq:xiLBd}
\xi^\mu &=& \hat{\lambda}_{1\nu} (\tilde{\partial}^\nu \xi_2^\mu - \tilde{\partial}^\mu \xi_2^\nu  )-\hat{\lambda}_{2\nu} (\tilde{\partial}^\nu \xi_1^\mu - \tilde{\partial}^\mu \xi_1^\nu  ) + \tilde{\partial}^\mu (\xi_1 \hat{\lambda}_2 - \xi_2 \hat{\lambda}_1 ) \\ \notag
&& 2 B_{\nu \rho} (\xi_1^\nu \tilde{\partial}^\rho \xi_2^\mu - \xi_2^\nu \tilde{\partial}^\rho \xi_1^\mu ) + 2 \tilde{\partial}^\mu B_{\nu \rho} \xi_1^\nu \xi_2^\rho
\, ,\\ \notag 
\hat{\lambda}_\mu &=&\hat{\lambda}_{1\nu} \tilde{\partial}^\nu \hat{\lambda}_{2\mu} -  \hat{\lambda}_{2\nu} \tilde{\partial}^\nu \hat{\lambda}_{1\mu} \\ \notag
&&- 2B_{\mu \nu} \Big( \hat{\lambda}_{1\rho} (\tilde{\partial}^\nu \xi^\rho_2- \tilde{\partial}^\rho \xi^\nu_2)- \hat{\lambda}_{2\rho} (\tilde{\partial}^\nu \xi^\rho_1- \tilde{\partial}^\rho \xi^\nu_1)- \frac{1}{2} \tilde{\partial}^\nu (\hat{\lambda}_1 \xi_2 -\hat{\lambda}_2 \xi_1)\Big) \\ \notag
&& + 2 \hat{\lambda}_{1\nu} \tilde{\partial}^\nu (\xi^\rho_2 B_{\rho \mu})- 2 \hat{\lambda}_{2\nu} \tilde{\partial}^\nu (\xi^\rho_1 B_{\rho \mu})+ 2(\xi_1^\nu B_{\nu \rho}) \tilde{\partial}^\rho \hat{\lambda}_{2\mu}-2(\xi_2^\nu B_{\nu \rho}) \tilde{\partial}^\rho \hat{\lambda}_{1\mu} \\ \notag
&& + 4\Big( B_{\mu \sigma} \tilde{\partial}^\sigma B_{\nu \rho} +B_{\nu \sigma} \tilde{\partial}^\sigma B_{\rho \mu}+B_{\rho \sigma} \tilde{\partial}^\sigma B_{\mu \nu} \Big) \xi_1^\nu \xi_2^\rho \, .
\end{eqnarray} 
In the T-dual description, $B_{\mu \nu}(y)$ plays the role of the T-dual bi-vector $B_{\mu \nu} =\frac{\kappa}{2} {^\star \theta}_{\mu \nu}$, while the parameters of general coordinate and local gauge transformations correspond to the parameters of local gauge and general coordinate transformations of the initial theory, respectively \cite{dualsim, DFT1}
\begin{equation}
{^\star \hat{\lambda}}^\mu = \xi^\mu \, , \quad {^\star \xi}_\mu = \hat{\lambda}_{\mu} \, . 
\end{equation}
As such, the bracket defined by (\ref{eq:xiLBd}) is ${^\star \theta}$-twisted Courant bracket.

Similarly, for the derivatives appearing in the $\theta$-twisted $C$-bracket we have
\begin{equation}
\breve{\partial}^Q \to \begin{pmatrix}
\delta^\mu_\nu & \kappa \theta^{\mu \nu} \\
0 & \delta^\nu_\mu 
\end{pmatrix} 
\begin{pmatrix}
\tilde{\partial}^{\nu} \\
0
\end{pmatrix} =
\begin{pmatrix}
\tilde{\partial}^{\mu}\\
0
\end{pmatrix} \, .
\end{equation}
Next, we write
\begin{equation} \label{eq:ctth1}
\breve{\Lambda}_1^N     \breve{\partial}_N  \breve{\Lambda}_2^Q \to \begin{pmatrix}
 \lambda_{1\nu}  \tilde{\partial}^{\nu} \breve{\xi}_2^{\mu}  \\ 
 \lambda_{1\nu}  \tilde{\partial}^{\nu} \lambda_{2 \mu}  
\end{pmatrix}
\end{equation}
and 
\begin{equation}\label{eq:ctth2}
\eta_{M N} \breve{\Lambda}_1^M \breve{\partial}^Q \breve{\Lambda}_2^N \to 
\begin{pmatrix}
\lambda_{1\nu}  \tilde{\partial}^{\mu} \breve{\xi}_2^{\nu}  + \breve{\xi}_1^{\nu} \tilde{\partial}^{\mu} \lambda_{2\nu}  \\
0
\end{pmatrix}
 \, ,
\end{equation}
while the flux term is simply given by
\begin{equation} \label{eq:ctth3}
\breve{\Lambda}_{1 M}  \breve{\Lambda}_{2 N}   \breve{\Theta}^{M N Q} \to 
\begin{pmatrix}
\kappa \ {{}^\star B}^{\mu \nu \rho} \lambda_{1\nu } \lambda_{2\rho}  \\
0
\end{pmatrix}
 \, ,
\end{equation}
where ${{}^\star B}^{\mu \nu \rho}$ is the $H$ flux in T-dual theory
\begin{eqnarray} \label{eq:tildeTh}
\kappa {{}^\star B}^{\mu \nu \rho} &=& \kappa \ \tilde{\partial}^\mu \theta^{\nu \rho} +\kappa \ \tilde{\partial}^\nu \theta^{\rho \mu}  + \kappa \ \tilde{\partial}^\rho \theta^{\mu \nu } \\ \notag
&=& 2 \tilde{\partial}^\mu \ {{}^\star B}^{\nu \rho} +2 \tilde{\partial}^\nu  \ {{}^\star B}^{\rho \mu}  +2 \tilde{\partial}^\rho  \ {{}^\star B}^{\mu \nu } 
 \, .
\end{eqnarray}
The full bracket 
\begin{equation}
[ \breve{\Lambda}_1, \breve{\Lambda}_2 ]_{{\textbf C}_{\theta}} \to [ \breve{\Lambda}_1, \breve{\Lambda}_2 ]_{{\cal{C}}_{\theta}} = \breve{\Lambda} \equiv
\begin{pmatrix}
\breve{\xi}\\
\lambda
\end{pmatrix}  \, ,
\end{equation}
where
\begin{eqnarray} \label{eq:xiLB2}
\breve{\xi}^\mu &=& \lambda_{1\nu} (\tilde{\partial}^\nu \breve{\xi}^\mu_2 -\tilde{\partial}^\mu \breve{\xi}^\nu_2  )-\lambda_{2\nu} (\tilde{\partial}^\nu \breve{\xi}^\mu_1 -\tilde{\partial}^\mu \breve{\xi}^\nu_1  )+\frac{1}{2}\tilde{\partial}^\mu (\breve{\xi}_1 \lambda_2 - \breve{\xi}_2 \lambda_1) \\ \notag
&&+\kappa \ {{}^\star B}^{\mu \nu \rho}  \lambda_{1\nu } \lambda_{2\rho}  \, , \\ \notag
\lambda_\mu &=& \lambda_{1\nu} \tilde{\partial}^\nu \lambda_{2\mu} - \lambda_{2\nu} \tilde{\partial}^\nu \lambda_{1\mu} \, .
\end{eqnarray}
We recognize the bracket as the Courant bracket twisted by a 2-form ${}^\star B$, written in terms of T-dual variables. 

In the double theory we naturally obtain both the Courant bracket twisted by a 2-form $B$ and bi-vector $\theta$ from a single twisted $C$-bracket. These twisted Courant brackets define Courant algebroids that are mutually isomorphic, where the isomorphism connecting them represents the T-duality \cite{crdual, calgdual}.

\section{Conclusions}
\cleq

We started with the theory characterized solely by the metric tensor and considered firstly generator of diffeomorphisms and local gauge transformations, equivalent to the T-dual diffeomorphisms. This generator is parametrized by a double parameter, whose components depend on both the initial and T-dual coordinates. It has already been shown to give rise to the $C$-bracket, which is the double theory generalization of the well known Courant bracket. We followed the method of \cite{csd} to obtain the twisted $C$-brackets.

Primarily, we considered the twist by a 2-form $B_{\mu \nu} (x,y)$ and obtained the $C$ bracket twisted by $B$. The resulting bracket can be separated into two parts. The first part of the bracket has identical form as the $C$-bracket, though the derivatives $\hat{\partial}^M = \partial^M + \hat{B}^M_{\ N}\  \partial^N$ are found in place of the usual $\partial^M$ derivatives. The second part of the twisted $C$-bracket contains a generalized flux $\hat{B}^{MNQ}$ (\ref{eq:BFlux}). 

Secondarily, we obtained the $C$-bracket twisted by a bi-vector $\theta$, which also contains a part of the same form as the $C$-bracket, and the flux contracting two gauge parameters. Similarly to the previous case, the bracket is written in terms of different derivatives $\breve{\partial}^M = \partial^M + \hat{\theta}^M_{\ N} \ \partial^N $, while the generalized flux is $\breve{\Theta}^{MNQ}$ (\ref{eq:breveTh}). There are a couple of ways how the twisted $C$-brackets differ from their twisted Courant counterparts.

Firstly, we noted that the $B$-twisted (\ref{eq:C-B}) and $\theta$-twisted (\ref{eq:C-Th}) $C$-brackets have the same form, with only difference being that the derivatives and fluxes (\ref{eq:BBS}) of the former are given in terms of 2-form $B$, while of the latter in terms of bi-vector $\theta$ (\ref{eq:TTR}). As it can be easily seen from comparing (\ref{eq:xiLB}) to (\ref{eq:xiLR}), this is not the case for their analogous twisted Courant brackets. If we take into the account that the Kalb-Ramond field and the non-commutativity parameter are mutually T-dual, the T-duality between the $B$- and $\theta$-twisted $C$-brackets is obvious.

Secondly, the twisting of the Courant bracket can be realized by adding the terms with fluxes to the standard Courant bracket expression. When the $C$-bracket is twisted, apart from adding the fluxes, the derivatives also change and transform by the same twisting matrix. The derivatives transform in the same way as $O(D,D)$ vectors, and do not result in the derivatives in the new basis.

Thirdly, both of the twisted $C$-brackets separately encapsulate the isomoprhism between two Courant algebroids, which represents the T-duality. This was shown by considering the initial and T-dual theories separately, by neglecting the T-dual coordinate dependence in the former and the initial coordinate dependence in the latter case. We have shown that in the same way that $C$-bracket becomes the Courant bracket, the twisted $C$-brackets become their twisted Courant counterparts. Moreover, we showed that in the T-dual description, the $B$-twisted $C$-bracket becomes in fact $\theta$-twisted Courant bracket, and vice versa.

\end{document}